\documentclass{iaus}
\usepackage{graphicx,natbib,astro}
\title[Galactic Center Jet] 
{The Jet in the Galactic Center: An Ideal Laboratory for
  Magnetohydrodynamics and General Relativity}

\author[Falcke et al.]   
{Heino Falcke$^{1,2}$, 
Sera Markoff$^{3}$, 
Geoffrey C. Bower$^{4}$, 
Charles F. Gammie$^{5,6}$, 
Monika Mo\'scibrodzka$^5$, \and
Dipankar Maitra$^7$ 
}

\affiliation{$^1$Department of
  Astrophysics, Institute for Mathematics, Astrophysics and Particle
  Physics (IMAPP), Radboud University, Nijmegen, The Netherlands \\[\affilskip] 
$^2$ASTRON, Oude
  Hoogeveensedijk 4, 7991 PD Dwingeloo, The Netherlands \\[\affilskip]
$^3$Astronomical Institute ``Anton Pannekoek'', University of
  Amsterdam, The Netherlands \\[\affilskip]
$^4$ Astronomy Department \& Radio Astronomy Lab, UC Berkeley, USA\\[\affilskip]
$^5$Department of Physics, University of Illinois,
Urbana, Illinois, USA\\[\affilskip]
$^6$Astronomy Department, University of Illinois, Urbana, Illinois, USA \\[\affilskip]
$^7$Department of Astronomy, University of Michigan, Ann Arbor,
Michigan, USA 
}

\pubyear{2010}
\volume{275}  
\pagerange{1--9}
\jname{Jets at all Scales}
\editors{G.E. Romero, R.A. Sunyaev \& T. Belloni, eds.}
\begin{document}

\maketitle

\begin{abstract}
  Of all possible black hole sources, the event horizon of the
  Galactic Center black hole, Sgr A*, subtends the largest angular
  scale on the sky. It is therefore a prime candidate to study and
  image plasma processes in strong gravity and it even allows imaging
  of the shadow cast by the event horizon. Recent mm-wave VLBI and
  radio timing observations as well as numerical GRMHD simulations now
  have provided several breakthroughs that put Sgr A* back into the
  focus. Firstly, VLBI observations have now measured the intrinsic
  size of Sgr A* at multiple frequencies, where the highest frequency
  measurements have approached the scale of the black hole
  shadow. Moreover, measurements of the radio variability show a clear
  time lag between 22 GHz and 43 GHz. The combination of size and
  timing measurements, allows one to actually measure the flow speed
  and direction of magnetized plasma at some tens of Schwarzschild
  radii.  This data strongly support a moderately relativistic
  outflow, consistent with an accelerating jet model. This is compared
  to recent GRMHD simulation that show the presence of a moderately
  relativistic outflow coupled to an accretion flow Sgr A*. Further
  VLBI and timing observations coupled to simulations have the
  potential to map out the velocity profile from 5-40 Schwarzschild
  radii and to provide a first glimpse at the appearance of a jet-disk
  system near the event horizon. Future submm-VLBI experiments would
  even be able to directly image those processes in strong gravity and
  directly confirm the presence of an event horizon. 
\end{abstract}

\firstsection 
\section{Introduction}
Based on an analogy to other active galactic nuclei (AGN)
\cite{Lynden-BellRees1971} proposed to look for a compact radio source
in the center of our own Milky Way, which was then discovered by
\cite{BalickBrown1974} only a little later with the NRAO
interferometer at Green Bank -- barely beating the team at Westerbork
\citep{EkersGossSchwarz1975}. This ``compact radio source in the
Galactic Center'' became later known as ``Sagittarius A*'' (Sgr A*)
and by now is the best constrained super-massive black hole candidate
we know of to date \citep[see][for a
review]{MeliaFalcke2001,GenzelEisenhauerGillessen2010}, with a mass of
$4\times10^6 M_\odot$ measured from stellar orbits
\citep{SchodelOttGenzel2002,GhezSalimWeinberg2008,GillessenEisenhauerFritz2009}.
Sgr~A* serves as an excellent example to understand black holes (and
compact radio cores) at very low accretion rates and is a potentially
exciting laboratory for the study of General Relativity (GR)
itself. In fact, \cite{FalckeMeliaAgol2000} argued that, what they
called the ``shadow of the black hole event horizon'' could reasonably
be detected with mm-wave very long baseline interferometry (VLBI) in
this source and many papers have since illuminated various general
relativistic effects that could be tested in this way
\citep{BroderickLoeb2006,FishDoelemanBroderick2009,HarkoLobo2009,YuanCaoHuang2009,JohannsenPsaltis2010,DexterAgolFragile2010}.

However, ever since its discovery, a discussion raged about the exact
nature of its emission processes and the region where the observable
radiation is emitted. Already \cite{ReynoldsMcKee1980} suggested an
origin of the radio emission in a jet or wind from a stellar-sized
object. Indeed, the properties of Sgr A* resemble those of
flat-spectrum compact radio cores in quasars (later also found in
X-ray binaries), which were explained by \cite{BlandfordKonigl1979} as
the $\tau$=1 surfaces (optical depth becomes unity) of powerful 
relativistic jets. This issue was revisited by
\cite{FalckeMannheimBiermann1993,FalckeMarkoff2000} who explained
spectrum and size of Sgr A* as a scaled-down quasar jet from an ``AGN
on a starvation diet'', i.e. a supermassive black hole with very low
accretion rate ($\dot M<10^{-7} M_\odot/{\rm yr}$). This semi-analytic
approach was based based upon a modified Blandford-K\"onigl model
\cite[][taking the longitudinal pressure gradient into
account]{Falcke1996a} and the jet-disk symbiosis ansatz
\citep{FalckeBiermann1995,FalckeBiermann1999}, which introduced a very
simple, yet until to day very effective linear scaling between
jet-power and accretion disk rate for black holes $Q_{\rm jet}=q_{\rm
  j}\dot M_{\rm disk}$, where $q_{\rm j}\sim3-10\%.$ Based on the
jet-disk symbiosis model, a number of concrete predictions were made
\citep{Falcke1999b} that all have stood the test of time.

Alternatively, it was proposed that the radio emission -- and
subsequently the emission at other wavelengths -- was produced in the
accretion flow itself. While standard optically thick, geometrically
thin accretion models started to fail in explaining the low infrared
flux of Sgr A* \citep{FalckeMelia1997,CokerMeliaFalcke1999}, the picture of an
optically thin, geometrically thick accretion flow, so-called
advection-dominated or radiatively-inefficient accretion flows (ADAFs,
RIAFs) started to emerge
\citep{Melia1992a,NarayanMahadevanGrindlay1998,QuataertGruzinov2000a}.
These models initially had much higher accretion rates, but this was
brought down when polarization measurements, using Faraday rotation
arguments,
\citep{BowerBackerZhao1999,AitkenGreavesChrysostomou2000,BowerFalckeWright2005,MarroneMoranZhao2007}
showed a low particle density towards Sgr A*. This implied an accretion
rate below $10^{-7}M_\odot$/yr. Hence, our Galactic Center is likely
starving as well as radiating inefficiently and producing a jet
\citep{YuanMarkoffFalcke2002}.

\section{Review of Observational Progress}
In recent years, important progress has come from the detection
of Sgr A* in X-rays and near-infrared
\citep{BaganoffBautzBrandt2001,GenzelOttEckart2003} and a lot of
emphasis has been placed on variability studies of Sgr A* \cite[][to
name just a
few]{HerrnsteinZhaoBower2004,Falcke1999a,PorquetGrossoPredehl2008,MauerhanMorrisWalter2005,Yusef-ZadehRobertsWardle2006,MarroneBaganoffMorris2008,EckartBaganoffZamaninasab2008,MeyerDoGhez2008,Yusef-ZadehBushouseWardle2009,Dodds-EdenPorquetTrap2009}. The
various campaigns have shown that flares in the near-infrared (NIR)
and X-rays are simultaneous within minutes and hence come from the same
region. \cite{Dodds-EdenPorquetTrap2009} favor optically thin
synchrotron radiation over inverse Compton as the dominant emission
process to explain the remarkable similarity of a particularly bright
flare in both bands. On the other hand radio/submm-wave and X-ray/NIR
flares do not seem to be simultaneous and no generally accepted lag
has been derived, despite various claims in the literature, which are
ranging from several hours to minutes.

\begin{figure}[htb]
\begin{center}
 \includegraphics[width=0.49\textwidth]{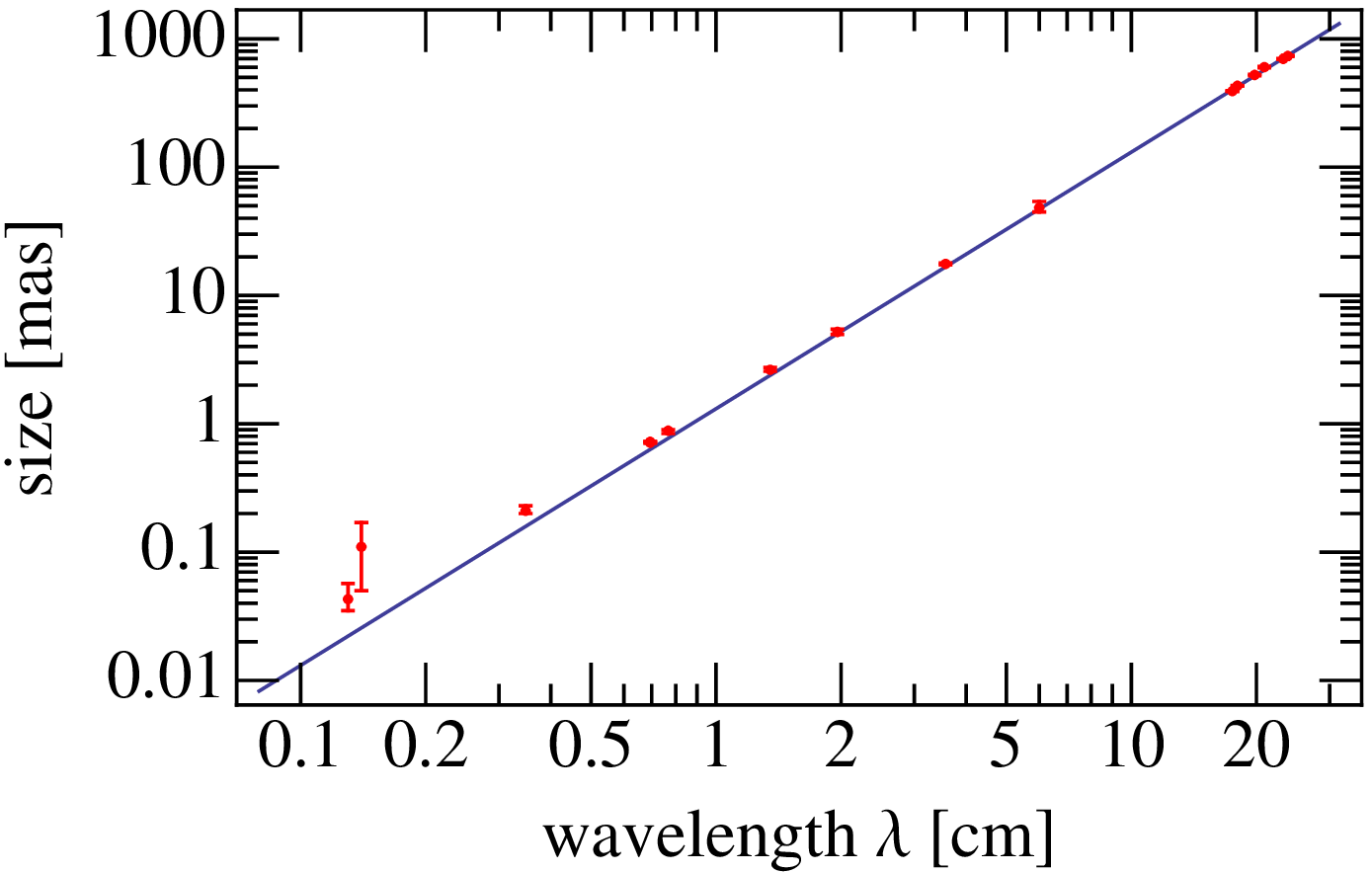} \hfill \includegraphics[width=0.49\textwidth]{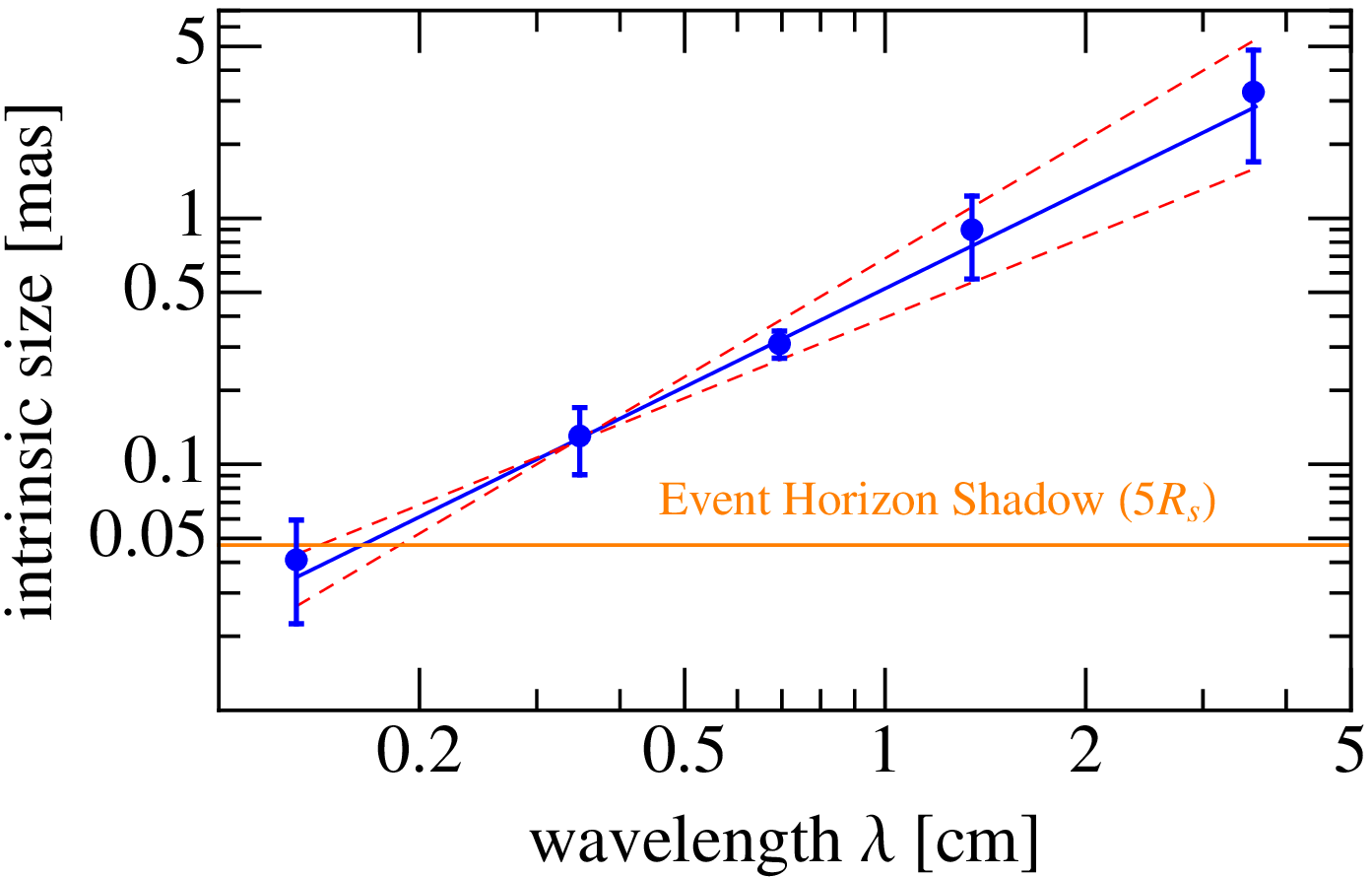} 
 \caption{Left: measured major axis size of Sgr A* as function of
   wavelength measured by various VLBI experiments. Right: Derived intrinsic size of Sgr A* after
   subtraction of a $\lambda^2$ scattering law using all available
   VLBI data (see
   \cite{FalckeMarkoffBower2009} for details). The dashed line
   indicates the systematic uncertainties due to different
   normalizations of the scattering law.}
   \label{figsgrsize}
\end{center}
\end{figure}

Very exciting has also been the progress based on radio observations,
constraining any model -- and in particular any jet model -- much
better. \cite{BowerFalckeHerrnstein2004} were the first to measure the
{\it intrinsic} source size of Sgr A* at 43 GHz directly using Very
Long Baseline Interferometry (VLBI), yielding a size of some 20
Schwarzschild radii ($R_{\rm s}$) only. This was a true breakthrough,
given that the structure of Sgr A* had been washed out due to
interstellar scattering for 30 years since its discovery (yielding a
measured size that decreases with $\lambda^2$ at low frequencies,
Fig.\,\ref{figsgrsize}, left).  Further VLBI measurements then
provided measurements up to 220 GHz
\citep{ShenLoLiang2005,DoelemanWeintroubRogers2008} giving sizes down
to $4 R_{\rm s}$! Hence, it is clear now that the intrinsic source
size decreases with increasing frequency (Fig.\,\ref{figsgrsize},
right) -- an important prediction of the jet model. Direct comparison
of the jet model with the actual VLBI data shows a consistency with
the compact source size and structure, but does favor edge-on
geometries \citep{MarkoffBowerFalcke2007}.

Additional information has come from the measurement of time lags in
the radio
\citep{Yusef-ZadehRobertsWardle2006,Yusef-ZadehWardleHeinke2008}, where
it was shown that 43 GHz radio flares precede 22 GHz flares by about
20 minutes. Given that the intrinsic size difference between these two
frequencies is about 30 light minutes one arrives at the natural
conclusion that the radio emitting plasma flows out with the speed of
light \citep{FalckeMarkoffBower2009}, which is now also suggested by the
short-time variability of the source
\citep{Yusef-ZadehBushouseWardle2009}. 

Alternative models based on simple adiabatic expansion of ``blobs''
which expand sub-relativisitically
\citep{Yusef-ZadehRobertsWardle2006,EckartBaganoffMorris2009} have
only been fitted to the light curves and fail to take the VLBI sizes
into account. They predict much smaller source sizes and would
postulate two very different emission regions, while fits for
adiabatic expansion in a jet \citep{MaitraMarkoffFalcke2009} do
reproduce radio light curves and VLBI sizes very well.

In the following we will now focus on the the basic properties of
emission from a jet and briefly discuss how future radio observations
and modeling will provide us with deeper insight into the formation of
jets close to the event horizon.

\section{The Basic Jet Model}
\begin{figure}[htb]
\begin{center}
 \hfill \includegraphics[width=0.99\textwidth]{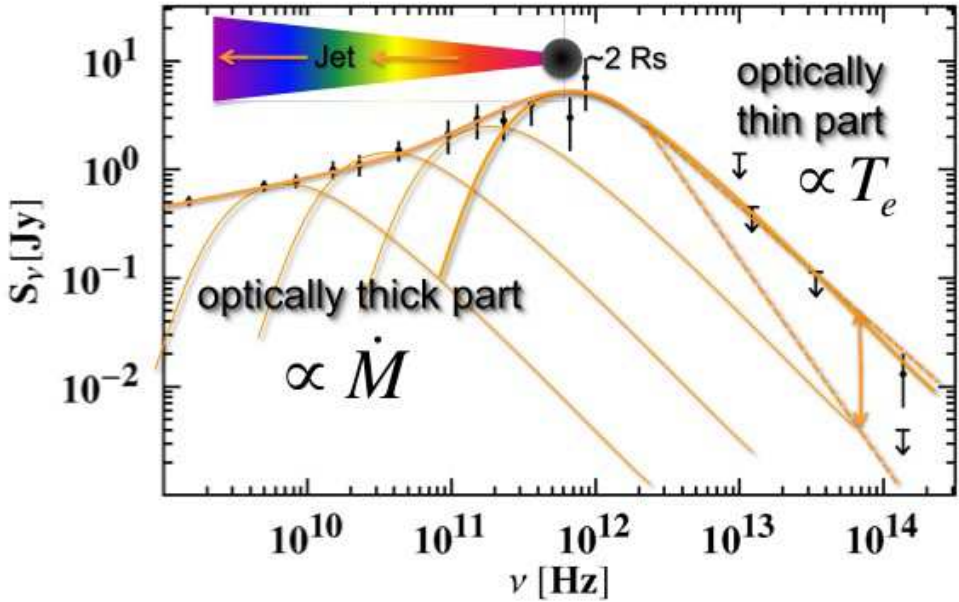} \hfill 
 \caption{Spectrum of Sgr A* and schematic overview of various
   contributions from a jet.}
   \label{figspectrum}
\end{center}
\end{figure}

Given the observational progress any model for Sgr A* nowadays has to
explain the spectrum, size, and variability properties of the
source. Here we will briefly explain the physical background, how many
of the observed properties arise quite naturally in a jet
model. Figure \ref{figspectrum} shows the radio spectrum of Sgr A* and
a schematic view of the various jet model components.  Assume a hot
magnetized plasma is ejected close to the black hole from a compact
nozzle and expands freely thereafter. The plasma in the nozzle will
emit synchrotron radiation peaking between 100 GHz and a few THz due
to optical depth effects (almost black body radiation), producing the
``submm-bump''. If the electron distribution is non-Maxwellian, i.e.,
has a high-energy power-law tail, optically thin synchrotron radiation
will appear, that can easily extend into the near-infrared or even
X-ray regime (which could also have a contribution from inverse
Compton emission from the submm-bump).

When the plasma leaves the nozzle and expands, it will expand with
sound speed and advance with the bulk flow speed, roughly filling a
conical Mach cone.  Like in a multi-color accretion disk, the
synchrotron emission will then peak at lower and lower frequencies the
further out the plasma gets.  This will naturally lead to an optically
thick, flat radio spectrum and a source size $\propto\nu^{-1}$
\citep{BlandfordKonigl1979,FalckeBiermann1995}. With the jet-disk
scaling mentioned above one can then readily explain flux and size of
Sgr A* to zeroth order \citep[see][for a text-book level
description]{Falcke2003}. However, a simple Blandford-K\"onigl model
is not really appropriate, since close to the black hole the pressure
gradient -- and likely other processes as well -- will lead to an
acceleration of the jet, yielding an inverted spectrum and a steeper
than $\nu^{-1}$ frequency-size relation
\citep{Falcke1996a,FalckeMarkoff2000}.

To understand the variability behavior one has to remember that the
radio/submm spectrum (<1 THz) is optically thick and thus originates
in a superposition of emission components from different spatial
regions, peaking at increasingly lower frequencies as their
characteristic scale increases.  The NIR/X-ray spectrum, on the other
hand, is optically thin and comes from one, i.e., the smallest,
spatial scale. Already \cite{MarkoffFalckeYuan2001} showed that the
jet radio emission will mainly respond to changes in the density (and
hence accretion rate, given the assumed jet-disk coupling), while the
NIR/X-ray emission will be highly sensitive to changes in the electron
temperature or power-law distribution. Comparing Sgr A* to a lightning
storm on earth, one might say that the radio would reflect the bulk
motion of the clouds and the winds, while NIR/X-rays are more like
lightning within the clouds.

This explains the high NIR/X-ray variability and its tight correlation
-- NIR and X-ray both reflect co-spatial changes in the population of
high-energy electrons in the nozzle region, close to the event
horizon. Compared to the bulk density of the flow, the number of
high-energy electrons is small and their cooling time is fast. Hence,
already a small change in the efficiency of accelerating non-thermal
electrons can lead to large changes in NIR and X-rays without
affecting the overall energy budget too much.

On the other hand, to produce radio flares a major change in
the bulk density of low-energy electrons is required, i.e., a change in
accretion rate and total jet power. When induced at the foot point the
density surge will have to propagate with the flow speed outwards
\citep[see][for detailed
modeling]{FalckeMarkoffBower2009,MaitraMarkoffFalcke2009}. This
suggests lower relative flux changes compared to NIR, time lags in the
sense that higher-frequencies lead lower ones, and not necessarily a
tight connection between major radio/mm--flares on one hand and
NIR/X-ray--flares on the other. Of course, it is conceivable that a
sudden increase in accretion rate onto the black hole, which is seen
as a radio flare, will also lead to more efficient dissipation and
particle acceleration. However, in any case NIR/radio delays are
likely not easy to interpret and should certainly not be used to
estimate, e.g., time scales for adiabatic expansion. On the other
hand, time lags within the radio/submm-wave regime could have high
diagnostic powers, since they are directly related to the flow
speed. Hence, VLBI size and radio monitoring data combined would
provide unique information about the acceleration and formation of
astrophysical jets and in particular the longitudinal velocity profile.

\section{MHD Modeling and Event Horizon Imaging}
Parallel to the semi-analytic modeling a new breed of MHD models has
become available, some of which are now geared towards Sgr A*. These
models basically start from first principles following the evolution
of a hot magnetized plasma as it is accreted onto a black
hole. Initially, jets did not form naturally in these simulations
\citep[][]{StoneHawleyGammie1996,Hawley2000}, especially when the
initial configuration contained only a toroidal magnetic field, while
simulations with random or poloidal fields seemed well to be able to
produce jet-like outflows
\citep[e.g.,][]{KoideShibataKudoh1999,MeierKoideUchida2001}. Nowadays,
most MHD simulations include poloidal magnetic field and hence show
some evidence for jet-like outflows. Given the excellent observational
constraints on scales close to the event horizon, Sgr A* has now
become a useful testbed also for numerical MHD simulations
\cite[e.g.,][]{OhsugaKatoMineshige2005,NobleLeungGammie2007,ChanLiuFryer2009,MoscibrodzkaGammieDolence2009,DexterAgolFragile2010,HilburnLiangLiu2010}.

For example, \cite{MoscibrodzkaGammieDolence2009} have performed
simulations based on an axisymmetric version of the GRMHD code {\it
  harm}, applying relativistic radiative transfer \citep[grmonty:
][]{DolenceGammieMoscibrodzka2009} and GR ray tracing \cite[ibothros:
][]{NobleLeungGammie2007} during post-processing. These 2D simulations
do indeed also show a jet-like outflow, but it appears rather
faint. Compared to the early pure GR ray tracing simulations
\citep{FalckeMeliaAgol2000} the shadow of the black hole remains
visible and a significant part of the parameter range of the models
can be excluded. In their simulations,
\cite{MoscibrodzkaGammieDolence2009} find that the VLBI sizes and
other properties favor edge-on orientations and a fast spinning black
hole.

\begin{figure}[htb]
\begin{center}\includegraphics[width=0.7\textwidth]{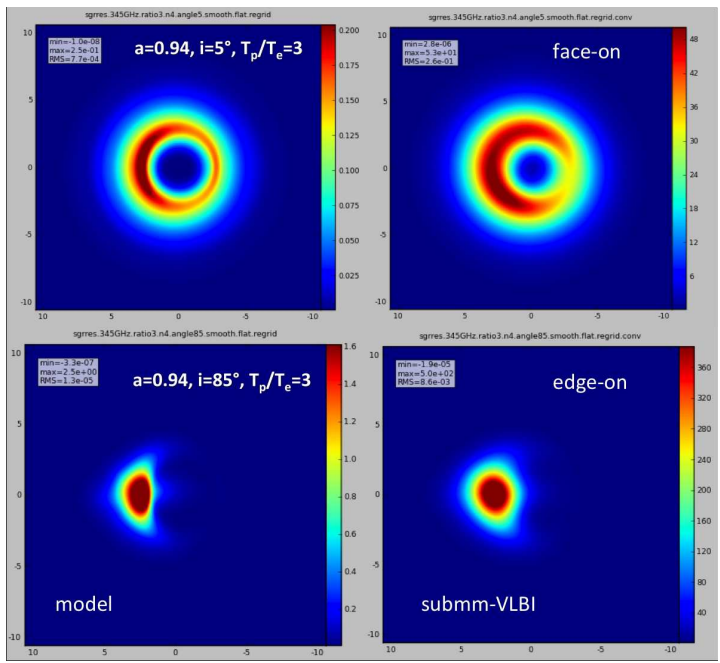}
 \caption{Left: Time-averaged appearance of 2D MHD accretion models at 345 GHz
   (bottom-left: best-fit model) from
   \cite{MoscibrodzkaGammieDolence2009}. Right: Reconstructed images
   with a putative global submm-VLBI array.  Top: face-on geometry,
   Bottom: edge-on geometry. The black hole shadow is clearly visible
   for face-one orientation but much less so for edge-on.}
   \label{figmhdvlbi}
\end{center}
\end{figure}

\begin{figure}[htb]
\begin{center}
\begin{tabular}{cc}
\includegraphics[width=0.4\textwidth]{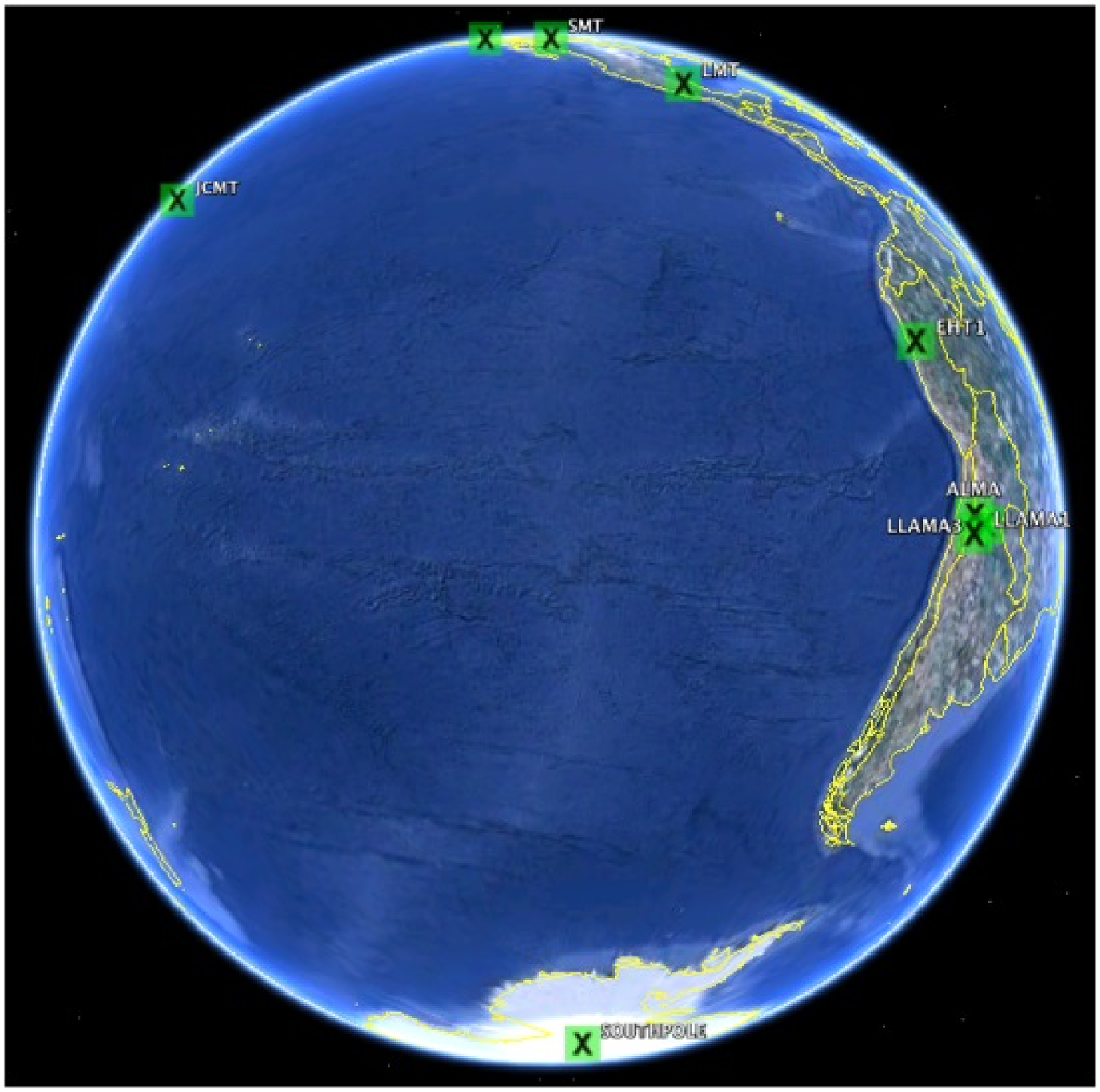} & \raisebox{0.2cm}{\includegraphics[width=0.394\textwidth]{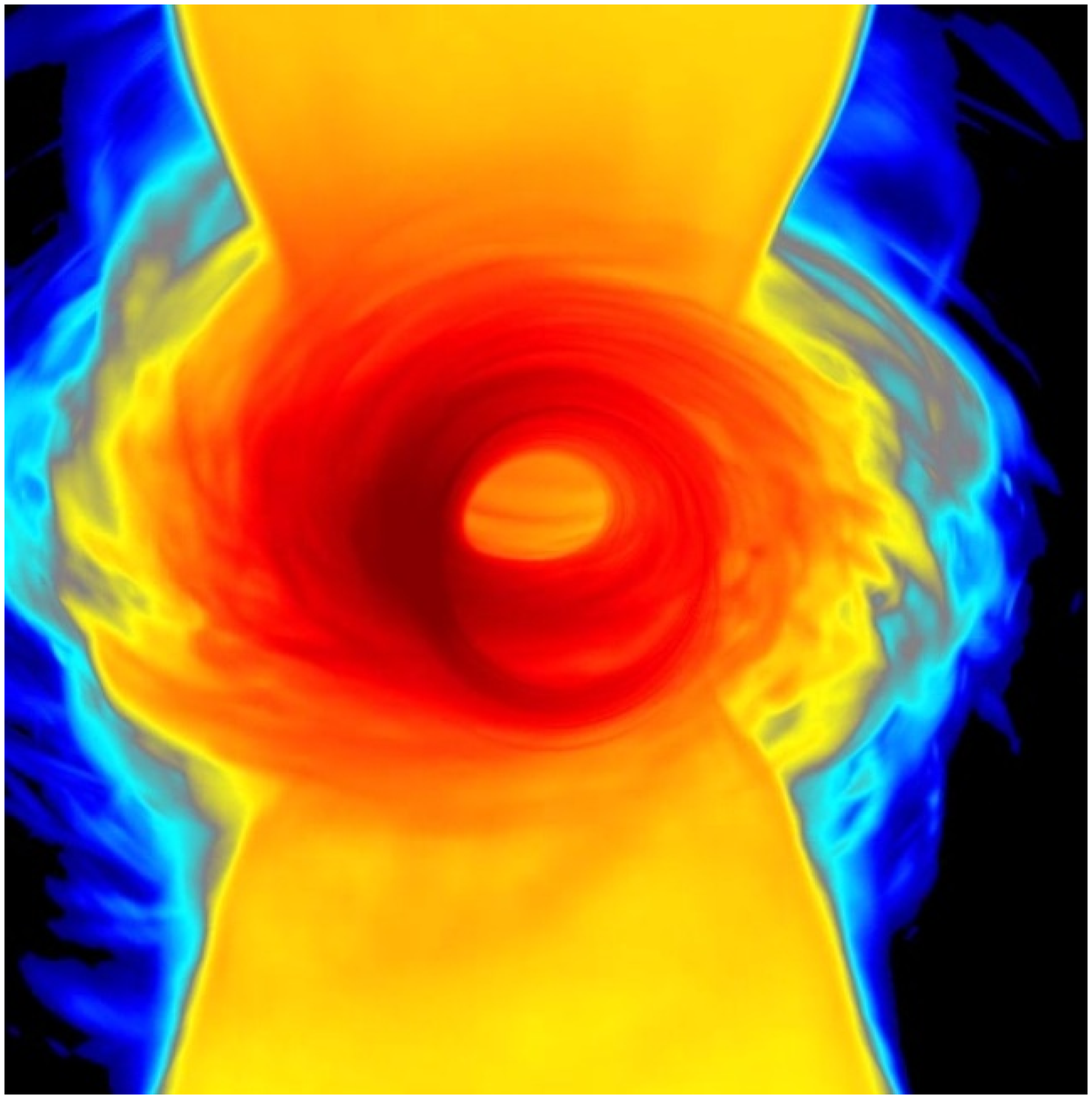}}
\end{tabular}
\caption{Left: Google-Earth view of the Earth with positions of the
  submm-VLBI stations used in the simulation as green boxes. Right:
  Snapshot of a 3D GRMHD simulation with similar parameters as in
  Fig.~\ref{figmhdvlbi}. Shown is the intensity at 345 GHz. The jet is
  much more visible, compared to the 2D case.}
   \label{figvlbilayout}
\end{center}
\end{figure}

To demonstrate the potential of this approach, we have performed some
simple simulations of future VLBI observations (Figure
\ref{figmhdvlbi}). For this we have taken the predicted intensity
distributions at 345 GHz (Fig.~\ref{figmhdvlbi}) from
\cite{MoscibrodzkaGammieDolence2009}, smeared them with a Gaussian
kernel according to the scattering law \citep[taken
from][]{FalckeMarkoffBower2009}, and passed them through the
interferometry simulation tool {\it simdata2} in {\it CASA}
\citep{McMullinWatersSchiebel2007}. Frequencies and baselines were
appropriately scaled to avoid numerical problems and noise
contributions were neglected. The array layout
(Fig.~\ref{figvlbilayout}) consisted of a number of existing
(sub)mm-wave facilities (CARMA, SMT, JCMT), telescopes which are
planned, are under construction or are not yet equipped for VLBI
(LLAMA, ALMA, LMT, SPT), and one putative site in Peru to fill in
missing baselines.

One can see that indeed such an array would be well matched to the
resolution needed to study event-horizon scale structures in Sgr
A*. High spin and inclination will produce a rather compact emission
region with a size of $\sim2-4 R_{\rm g}$ and a shadow that is difficult
to see unless one achieves sufficient dynamic range ($\sim100:1$).

It will be interesting to see how this picture changes as the
simulations improve. Figure~\ref{figvlbilayout} (right) shows a
snapshot of a single frame from a preliminary 3D GRMHD simulation with
{\it harm}, again including ray tracing and radiation transport, for a
black hole spin of a=0.9. While in the 2D simulation the jet is very
dim, to the point that it is not even visible in the emission maps
presented here, the jet becomes much more prominent in the 3D
case. This is apparently due to an increased electron temperature in
the jet, possibly coming from increased acoustic heating. Clearly, the
electron temperature in disk and jet plays and important role for the
appearance of Sgr A*, but it is also the least understood
parameter. Such simulations provide a taste for the future direction
the field can evolve to.

\section{Conclusions}
There is no other supermassive black hole with such a wealth of
information. The properties are not different from many other radio
cores in nearby low-power AGN when appropriately scaled to a very low
accretion rate \citep{FalckeBiermann1999,NagarFalckeWilson2005,MarkoffNowakYoung2008}. In
fact, within the fundamental plane of black hole activity
\citep{MerloniHeinzdiMatteo2003,FalckeKordingMarkoff2004,KordingFalckeCorbel2006},
that has been discussed at length in this meeting, it represents
almost an extreme ``off-state''.

Spectrum, size, and radio time-lags and evolution in Sgr A* strongly
suggest a jet-nature of the radio emission, with mildly relativistic
outflow speeds. Since the the submm-wave emission comes from the
direct proximity of the event horizon, Sgr A* is an ideal source to
image the shadow of the event horizon {\it and} to find out how
astrophysical jets are formed -- for example by combining mm-VLBI
observations and time-lag measurements.  
\bibliographystyle{aa}
\bibliography{hfrefs}

\begin{discussion}

\discuss{C. Fendt}{What is your heating model for the electrons. This
  is beyond MHD.}

\discuss{H. Falcke}{It is essentially local numerical dissipation.}

\discuss{Luis. F. Rodriguez}{Is there any chance of seeing the jet a
  few arcsec away from Sgr A* using an ultra-sensitive radio array like
the EVLA?}

\discuss{H. Falcke}{Unfortunately does the Sgr A* size grow roughly
  with $\lambda$ while the scattering size grows with $\lambda^2$, hence
  the core will always be hidden at lower frequencies. Perhaps one
  could see some interaction region further out, but teh Galactic Center is a very complicated
  region.}

\discuss{S. Corbel}{Do you have any idea why we live in a galaxy with
  the faintest LLAGN?}

\discuss{H. Falcke}{Estimates for the accretion rate based on
 environmental factors are all many orders of magnitude above the
  current value. Maybe we live in a special time. For example, the
  nearby supernova remnant Sgr A East has probably just swept over the
black hole a few thousand years ago. Such local effects could easily disrupt
a steady accretion flow.}

\discuss{E. de Gouveia Dal Pino}{Do you think that the time lag
  technique will be able to separate the complex motions that may
  occur at launching like, for instance, rotational velocity from the
  $v_{\rm z}$ component?}
\discuss{H. Falcke}{Good question -- helical motion is indeed quite
  commonly observed in quasar jets with VLBI. Rotation will lead to
  differential beaming and hence will imprint a wobble as a function
  of frequency on top of the flare spectrum. Maybe with very sensitive
  broad-band monitoring, e.g. with ALMA and SKA in the future, might
  such subtleties be detectable.}
\end{discussion}

\end{document}